\documentclass[a4paper,conference]{IEEEtran}
\usepackage{hyperref}
\usepackage{graphics} 
\usepackage{epsfig}
\usepackage{amssymb,amsmath}
\usepackage{lscape}
\usepackage{soul}
\DeclareMathAlphabet\mathbfcal{OMS}{cmsy}{b}{n}
\def\BibTeX{{\rm B\kern-.05em{\sc i\kern-.025em b}\kern-.08em
		T\kern-.1667em\lower.7ex\hbox{E}\kern-.125emX}}
\begin{document}
\title{Deep Learning for Screening  COVID-19 using Chest X-Ray Images}

\author{\IEEEauthorblockN{Sanhita~Basu}
	\IEEEauthorblockA{\textit{Department of Computer Science} \\
		\textit{West Bengal State University}\\
		West Bengal 700126, India\\
		tania.sanhita@gmail.com}
	\and
	\IEEEauthorblockN{Sushmita~Mitra}
	\IEEEauthorblockA{\textit{Machine Intelligence Unit} \\
		\textit{Indian Statistical Institute}\\
		Kolkata 700108, India\\
		sushmita@isical.ac.in}
	\and
	\IEEEauthorblockN{Nilanjan~Saha}
	\IEEEauthorblockA{\textit{Centre for Translational \& Clinical Research} \\
		\textit{Jamia Hamdard (Deemed University), }\\
		New Delhi 110062, India \\
		nilanjan.saha@jamiahamdard.ac.in}
}

\maketitle

\begin{abstract}
With the ever increasing demand for screening millions of prospective ``novel coronavirus'' or COVID-19 cases, and due to the emergence of high false negatives in the commonly used PCR tests, the necessity for probing an alternative simple screening mechanism of COVID-19 using radiological images (like chest X-Rays) assumes importance. In this scenario, machine learning (ML) and deep learning (DL) offer fast, automated, effective strategies to detect abnormalities and extract key features of the altered lung parenchyma, which may be related to specific signatures of the COVID-19 virus. However, the available COVID-19 datasets are inadequate to train deep neural networks. Therefore, we propose a new concept called domain extension transfer learning (DETL). We employ DETL, with pre-trained deep convolutional neural network, on a related large chest X-Ray dataset that is tuned for classifying between four classes \textit{viz.} $normal$, $pneumonia$, $other\_disease$, and $Covid-19$. A 5-fold cross validation is performed to estimate the feasibility of using chest X-Rays to diagnose COVID-19. The initial results show promise, with the possibility of replication on bigger and more diverse data sets. The overall accuracy was measured as $90.13\% \pm 0.14$. In order to get an idea about the COVID-19 detection transparency, we employed the concept of Gradient Class Activation Map (Grad-CAM) for detecting the regions where the model paid more attention during the classification. This was found to strongly correlate with clinical findings, as validated by experts.
\end{abstract}


\section{Introduction}
\IEEEPARstart {T}{he} coronavirus (CoV) belongs to a large family of viruses that cause illness ranging from the common influenza to the more severe manifestations, such as the Middle East Respiratory Syndrome (MERS-CoV) and the Severe Acute Respiratory Syndrome (SARS-CoV). The novel coronavirus (nCoV) or COVID-19 is a new strain called SARS-CoV2, and not previously identified in humans. Although this outbreak had its start as an epidemic in Wuhan, China, today it has severely affected multiple countries around the world as a pandemic. There is currently no effective cure for this virus and there is an urgent need to increase global knowledge in its mechanisms of infection, lung parenchyma damage distribution and associated patterns; not only for the disease detection or diagnosis, but also to support the design of curative therapy. Artificial Intelligence (AI) as applied to radiomic features from thoracic imaging, through X-Ray and Computed Tomography (CT), along with other clinical, pathologic and genomic parameters, is expected to provide useful support  in this direction. 

A critical step in the fight against COVID-19 is the effective screening of infected patients, such that those infected can receive immediate treatment and care, as well as be isolated to mitigate the spread of the virus. The gold standard screening method currently used for detecting COVID-19 cases is PCR  testing, which is a very time-consuming, laborious, and complicated manual process with kits presently in short supply. Besides, the test is uncomfortable, invasive, and uses nasopharyngeal swabs. On the other hand, X-ray machines are widely available and scans are relatively low cost. 

The need of the hour is, therefore, a fast detection; and this becomes all the more important as days progress and the healthcare system gets overwhelmed by the deluge of patient data. The necessity of designing an automated computerized process becomes all the more evident. With this in mind, we propose to employ radiomics \cite{gillies2015radiomics} from imaging, using deep learning \cite{lecun2015deep},  for the purpose. 

Deep learning \cite{lecun2015deep} entails learning from raw data to automatically discover the representations needed for detection or classification. In the context of medical images, it directly uses pixel values of the images (instead of extracted or selected features) at the input; thereby, overcoming the manual errors caused by inaccurate segmentation and/or subsequent feature extraction. Convolutional neural networks (CNNs) constitute one of the popular models of deep learning. The breakthrough in CNNs came with the ImageNet competition in 2012 \cite{krizh12}, where the error rate was almost halved for object recognition. 

AI algorithms, along with clinical and radiomic \cite{gillies2015radiomics} features derived from Chest X-rays, are expected to be of huge help to undertake massive detection programs that could take place in any country with access to X-ray equipment, and aid in effective diagnosis of COVID-19.
In this scenario, machine learning (ML) and deep learning (DL) offer fast, automated, effective strategies to detect abnormalities and extract key features of the altered lung parenchyma, which may be related to specific signatures of the COVID-19 virus. However, the available COVID-19 datasets are inadequate to train deep neural networks. 

This paper outlines our research in designing a novel algorithm called Domain Extension Transfer Learning (DETL), based on the concept of transfer learning for alternative screening of COVID-19 using chest X-Rays. We employ DETL, with pre-trained deep convolutional neural network, on a related large chest X-Ray dataset which is tuned for classifying between four classes \textit{viz.} $normal$, $other\_disease$, $pneumonia$ and $Covid-19$. The concept of Gradient Class Activation Map (Grad-CAM) is employed for detecting characteristic features from X-ray images to aid in visually interpretative decision making in relation to COVID-19.

\section{Thoracic Imaging}

Thoracic imaging is of great value in the diagnosis of COVID-19, monitoring of therapeutic efficacy, and patient discharge assessment \cite{liang2020handbook}. While a high-resolution CT is highly preferable, the portable chest X-rays are helpful for dealing with critically ill patients who are immobile. Daily routine portable chest X-rays are recommended for critically ill patients. Given that CT is more reliable to check changes in the lungs, as compared to the high false negatives in the commonly used random tests like PCR; its importance in the context of COVID-19 becomes all the more evident. Conspicuous ground-glass opacity and multiple mottling lesions, in the peripheral and posterior lungs on X-Ray and CT images, are indicative of COVID-19 pneumonia \cite{gozes2020rapid}. Therefore, it can play an important role in the diagnosis of COVID-19 as an advanced imaging evidence once the findings are indicative of coronavirus. 

It may be noted that bacterial and viral pneumonia also need to be initially distinguished from COVID-induced pneumonia, with normalization of the acquired X-ray images through different machines assuming utmost importance. Bacterial pneumonia is usually lobar, {\it ie.}, confined to one or more lobes. It is characterized by inflammatory exudates within the intra-alveolar space, resulting in consolidation that affects a large and continuous area of the lobe of a lung. A viral pneumonia, on the other hand, is usually interstitial; showing up in CT as diffuse bronchopneumonia or interstitial pneumonia across several fissures and lobes \cite{gozes2020rapid}. It is characterized by patchy foci of consolidation scattered in one or more lobes of one or both lungs. Now the COVID-induced pneumonia is like viral pneumonia, but is usually evident in the peripheral and lower parts of the lung. 

Teams in China and the United States found that the lungs of patients with COVID-19 symptoms had certain visual hallmarks, such as ground-glass opacities—hazy darkened spots in the lung diffuse enough that they don’t block underlying blood vessels or lung structures—and areas of increased lung density called consolidation \cite{gozes2020rapid}. Those characteristics became more frequent and were more likely to spread across both lungs the longer a person was infected. For Coronavirus patients the RADLogics Inc. (www.radlogics.com/) system output quantitative opacity measurements and a visualization of the larger opacities in a slice-based “heat map” or a 3D volume display. A suggested “Corona score” measured the progression of patients over time.

In order to speed up the discovery of disease mechanisms, as the medical systems get overwhelmed by data worldwide, machine learning (ML) and deep learning can be effectively employed to detect abnormalities and extract textural features of the altered lung parenchyma to be related to specific signatures of the COVID-19 virus. A preliminary deep convolutional network has been proposed \cite{wang2020covid}, based on open source 5941 chest radiography images across 2839 patient cases from two open access data repositories. More data is needed to consolidate the global database.

\section{Material and Methods}

Here we summarize the datasets used and the methodology employed.

\subsection{Datasets used}

\begin{figure}[]
	\centering\includegraphics[width=1.0\linewidth]{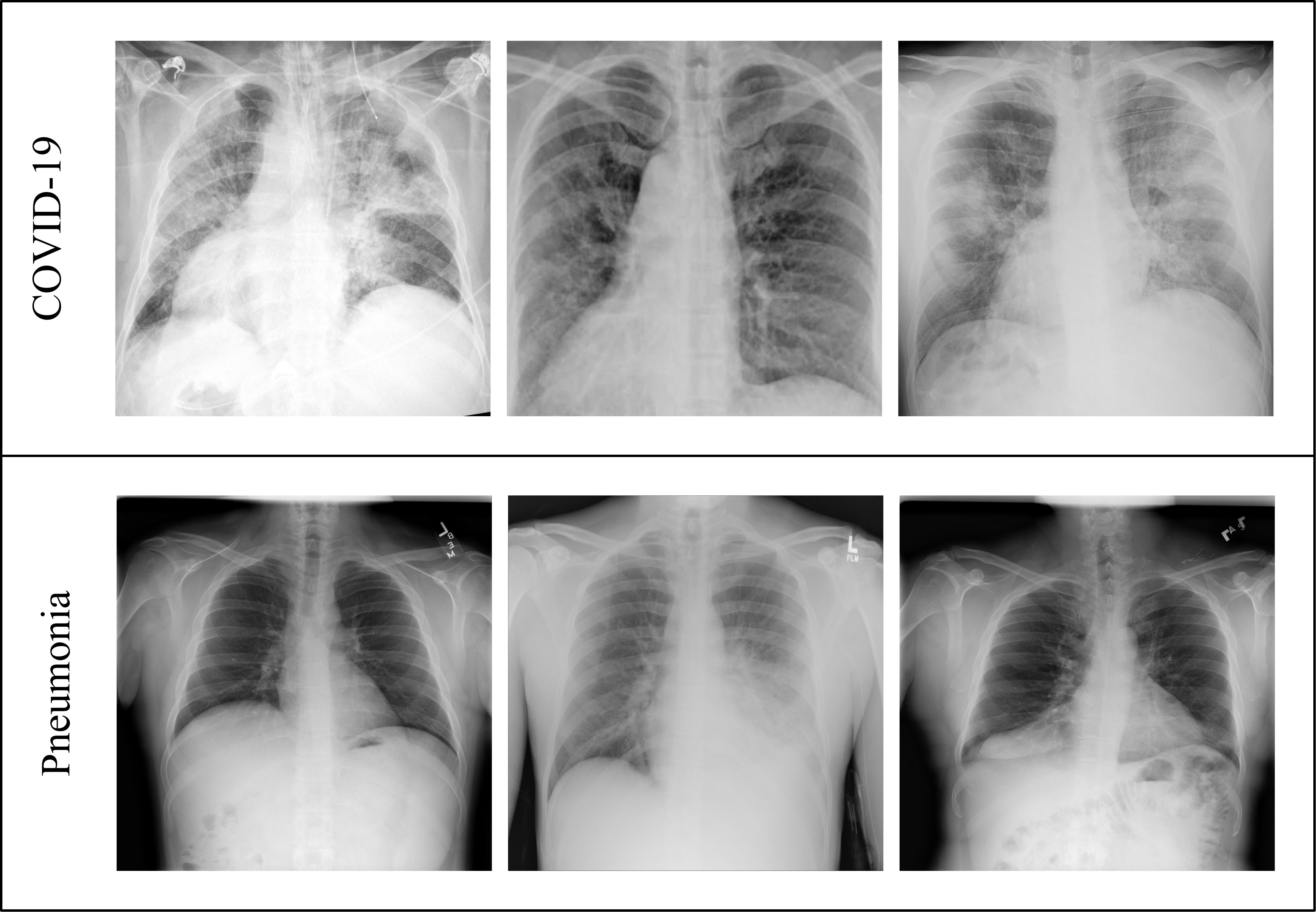}
	\caption{Sample X-Ray images of COVID-19 and Pneumonia.}
	\label{fig:fig1}
\end{figure}

\begin{figure}[]
	\centering\includegraphics[width=1.0\linewidth]{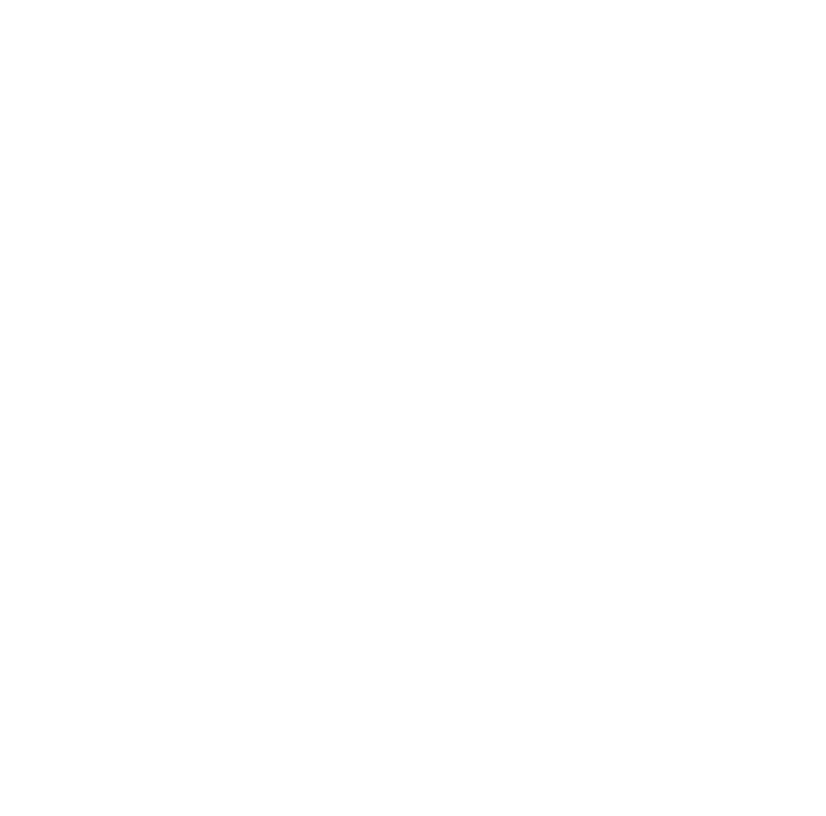}
	\caption{Co-occurrence statistics of 8 pathology classes, taken from Ref. \cite{wang2017chestx}.}
	\label{fig:fig2}
\end{figure}

A total of 305 COVID-19 X-Ray images were acquired from four open source databases, {\it viz.} (i) Italian Society of Medical Radiology and Interventional\footnote{\url{https://www.sirm.org/category/senza-categoria/covid-19/}} (25 cases), (ii) Radiopaedia.org (provided by Dr. Fabio Macori)\footnote{\url{https://radiopaedia.org/search?utf8=\%E2\%9C\%93&q=covid&scope=all&lang=us}} (20 cases), (iii) J. Paul Cohen {\it et al.} \cite{cohen2020covid} COVID-19 image data collection\footnote{\url{https://github.com/ieee8023/covid-chestxray-dataset}} (180 cases), and (iv) from a hospital in Spain (80 cases) \footnote{\url{https://twitter.com/ChestImaging/status/1243928581983670272}}. The chest X-Ray images of normal samples, and of 14 lung, heart and chest-related disease samples\footnote{Atelectasis, Consolidation, Infiltration, Pneumothorax, Edema, Emphysema, Fibrosis, Effusion, Pneumonia, Pleural thickening, Cardiomegaly, Nodule, Mass, and Hernia.} were obtained from the NIH Chest X-ray Dataset\footnote{\url{https://www.kaggle.com/nih-chest-xrays/data}}; composed of 108,948 frontal-view chest X-Ray images from 32,717 unique patients. Sample images of COVID-19 and Pneumonia are shown in Fig. \ref{fig:fig1}. Based on all these X-Ray data sources, we prepared two datasets termed Data-A and Data-B. The Data-A is generated from the ``NIH Chest X-ray Dataset'' and consists of two classes $normal$ and $disease$. The disease class is composed of all the 13 lung, heart and chest-related diseases, as mentioned above, excluding the pneumonia. Data-B has four classes, corresponding to $normal$, $other\_disease$, $pneumonia$ and $Covid-19$. Images for $normal$ and $pneumonia$ classes are taken from the ``NIH Chest X-ray Dataset''. While all the 322 images of the $pneumonia$ class were considered, the 350 images from the $normal$ class were randomly sampled. For the $other\_disease$ class we randomly picked 50 images from the six pathology classes, {\it viz.} ``Atelectasis'', ``Cardiomegaly'', ``Infiltration'', ``Effusion``, ``Nodule'' and ``Mass'', which sometimes co-occurred with pneumonia as depicted in Fig. \ref{fig:fig2} \cite{wang2017chestx}. It basically represents the co-occurrence of multiple pathology classes, which also matches with experts’ domain knowledge. For example, as evident from the figure, the class ``Infiltration'' is often associated with ``Atelectasis'' and ``Effusion''. Table \ref{tab:data} summarizes information related to the images in each of the classes of Data-A and Data-B. 

\begin{table}[]
\caption{Number of images in each of the classes of Data-A and Data-B.}
\begin{center}\label{tab:data}
\begin{tabular}{|ccc|}
\hline
\multicolumn{3}{|c|}{\textbf{Data-A}}                                       \\ \hline
                                 & $normal$                     & $diseased$ \\ \hline
\multicolumn{1}{|c|}{Training}   & \multicolumn{1}{c|}{$47560$} & $40819$   \\ \hline
\multicolumn{1}{|c|}{Validation} & \multicolumn{1}{c|}{$10000$} & $10000$   \\ \hline 
\end{tabular}

\vspace{0.23cm}

\begin{tabular}{|c|c|c|c|}
\hline
\multicolumn{4}{|c|}{\textbf{Data-B}}                  \\ \hline
$normal$ & $pneumonia$ & $other\_disease$ & $Covid-19$ \\ \hline
$350$    & $322$       & $300$            & $305$      \\ \hline
\end{tabular}
\end{center}
\end{table}

\subsection{Methods}

Often training a CNN from scratch is generally difficult because it essentially requires large training data, along with significant expertise, to select an appropriate model architecture for proper convergence. In medical applications data is typically scarce, and expert annotation is expensive. Training a deep CNN requires huge computational and memory resources, thereby making it extremely time-consuming. Transfer learning (TL) offers a promising alternative, in case of inadequate data, to fine tune a CNN already pre-trained on a large set of available labeled images from some other category. This helps in speeding up convergence while lowering computational complexity during training. Typically the early layers of a CNN learn low-level image features, which are applicable to most vision tasks. The later layers, on the other hand, learn high-level features which are more application-specific. Therefore, shallow fine-tuning of the last few layers is usually sufficient for transfer learning. A common practice is to replace the last fully-connected layer of the pre-trained CNN with a new fully-connected layer, having as many neurons as the number of classes in the new target application. The rest of the weights, in the remaining layers, of the pre-trained network are retained. This corresponds to training a linear classifier with the features generated in the preceding layer. 

The acquired COVID-19 dataset was considered as inadequate to train from scratch the CNN, involving a huge number of trainable parameters, to learn the complex representative features for distinguishing between COVID-19 and community acquired pneumonia images. Therefore TL was employed to leverage the knowledge (features, weights, etc.) learned from the source domain ($\mathcal{D}_S$) and source task ($\mathcal{T}_S$) for training newer models for the target domain ($\mathcal{D}_T$) and target task ($\mathcal{T}_T$) with limited annotated data. 

A domain ($\mathcal{D}$) can be represented by a tuple $\{\mathcal{X}, P(X)\}$, where $\mathcal{X}$ and $P(X)$ ($X = x_1, x_2, \dots, x_n \in \mathcal{X}$) represent the feature space and corresponding marginal probability distribution. Given a domain $\mathcal{D} = \{\mathcal{X}, P(X)\}$ a task $\mathcal{T}$ consists of a label space 
$Y$ and a conditional probability distribution $(Y|X)$ that is typically learned from the training data. Given source and target domains ($\mathcal{D}_S$, $\mathcal{D}_T$) and tasks ($\mathcal{T}_S$, $\mathcal{T}_T$), the objective of transfer learning is to learn the target conditional probability distribution $P(Y_T|X_T)$ in $\mathcal{D}_T$ with the knowledge gained from $\mathcal{D}_S$ by solving $\mathcal{T}_S$.

It is quite common in the literature to use models pre-trained on large datasets like ImageNet\footnote{\url{http://www.image-net.org/}} through TL for a new task involving a different dataset with limited data points. However, it has been recently established experimentally \cite{he2018rethinking} that if the source and target domains are very dissimilar in nature, such as in natural images and medical images, then TL has a very limited role to play as the networks may learn very different high-level features in the two cases. 
Therefore, in this research, we propose the concept of domain extension transfer learning where $\mathcal{D}_T  \supset \mathcal{D}_S$. 

Our approach consists of training a CNN model from scratch on Data-A, to learn to classify between diseased and normal X-ray images. Next we replace the last fully-connected layer of the pre-trained CNN with a new fully-connected layer, having as many neurons as the number of
classes; which, in our case, is four {\it viz.} $normal$, $other\_disease$, $pneumonia$ and $Covid-19$. The rest of the weights, in the remaining layers of the pre-trained network, are retained. This corresponds to training a linear classifier with the features generated in the preceding layer. The adopted procedure is outlined below.
\begin{itemize}
    \item Instantiate the convolutional base of the model trained on Data-A and load its pre-trained weights.
    \item Replace the last fully-connected layer of the pre-trained CNN with a new fully-connected layer.
    \item Freeze the layers of the model up to the last convolutional block.
    \item Finally retrain the last convolution block and the fully-connected layers using Stochastic Gradient Descent (SGD) optimization algorithm with a very slow learning rate.
\end{itemize}

Three popular CNN models, with increasing number of layers {\it viz.} AlexNet \cite{krizhevsky2012imagenet} (8 layers), VGGNet \cite{simonyan2014very} (16 layers) and ResNet \cite{he2016deep} (50 layers), were used. Even though ResNet is deeper as compared to VGGNet, the model size of ResNet becomes substantially smaller due to the usage of global average pooling in lieu of fully connected layers. Instead of using large kernels, as in  AlexNet ($11\times11$), VGGNet and ResNet, we employed smaller kernels of size $3 \times 3$. It may be noted that smaller kernels produce  better regularization  due to the smaller number of trainable weights,  with the possibility  of constructing  deeper networks  without losing too much information in the layers \cite{Pereira2016}. Initially all models were trained from scratch using Data-A, such that they learned to  distinguish between diseased and normal X-ray images. Next the models were fine-tuned on Data-B, using the TL strategy (discussed above), to let them learn to classify between the four classes \textit{viz.} $normal$, $other\_disease$, $pneumonia$ and $Covid-19$.    

\section{Experimental Setup and Results}

CNN models were developed using TensorFlow\footnote{\url{https://www.tensorflow.org/}}, with a wrapper library Keras\footnote{\url{https://keras.io/}} in Python. The experiments were performed on a Dell Precision 7810 Tower with 2x Intel Xeon E5-2600 v3, totalling 12 cores, 256GB RAM, and NVIDIA Quadro K6000 GPU with 12GB VRAM. Adam optimization algorithm was used for hyperparameter optimization for training the CNNs from the scratch on Data-A with an initial learning rate $10^{-3}$, and decayed according to cosine annealing. Real time data augmentation was also used in terms of random rotation, scaling, and mirroring. A balanced validation dataset of size 20,000 was created by random sampling from the training set and used for validating the CNN model after each training epoch, for parameter selection and detection of overfitting. The model was trained for 100 epochs and the weights which achieved the best validation accuracy were retained. During TL of the pre-trained CNN on Data-B, we used SGD optimization algorithm with a learning rate $10^{-4}$ and momentum $0.9$. 

\begin{figure*}[]
	\centering\includegraphics[width=1.0\linewidth]{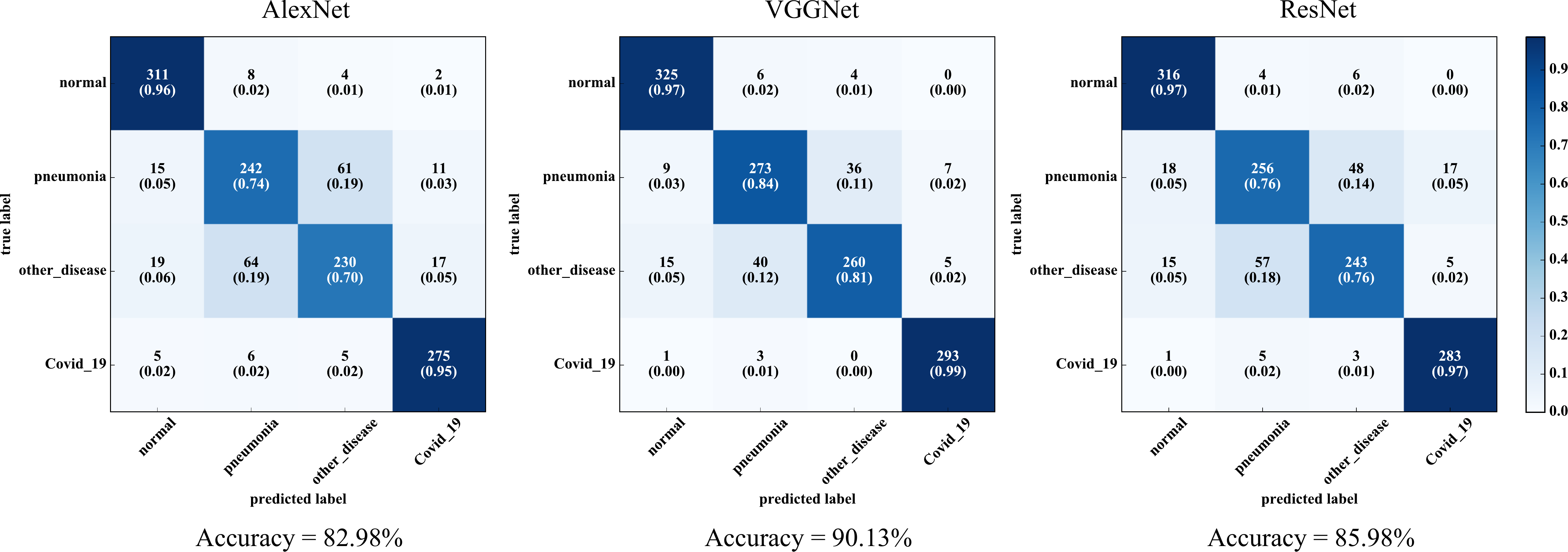}
	\caption{Confusion matrices for Alexnet, VGGNet and ResNet on Data-B.}
	\label{fig:fig4}
\end{figure*}

A 5-fold cross validation was performed to get an estimate of the feasibility of using chest X-Rays to diagnose COVID-19 with Data-B. The initial results show promise, subject to their possible replication on bigger and more diverse datasets. The 5-fold cross validation accuracy for AlexNet, VGGNet, and ResNet was measured as $82.98\% \pm 0.02$, $90.13\% \pm 0.14$, and $85.98\% \pm 0.07$, respectively. The sum the confusion matrices, generated over the different folds for the three CNNs, are shown in Fig. \ref{fig:fig4}. As observed, VGGNet was the best performing network. Although ResNet achieved the best validation accuracy on Data-A, the VGGNet correctly classified $99\%$ of the $Covid$ and $100\%$ of $normal$ cases in most of the validation folds of Data-B. There was some misclassification  between the $pneumonia$ and $other\_disease$ classes (as both are often  co-occurring).

\section{Decision Visualization}

In order to get an idea about the COVID-19 detection transparency, we employed the concept of Gradient Class Activation Map (Grad-CAM) \cite{selvaraju2017grad} for detecting the regions where the model paid more attention during the classification. Let $\mathcal{A}^i\in \mathbb{R}^{H \times W}$ represent the $i$th feature map ($i = 1, 2, \dots, K$) of the last convolution layer of the CNN. Here $K$ represents the total number of feature maps in the last convolution layer, with  $H$ and $W$ denoting the dimension of each feature map. Grad-CAM utilizes the $\mathcal{A}^i$'s to visualize the decision made by the CNN. The feature maps in the final convolution layer retain the spatial information that captures the visual pattern used for discriminating between the different classes. 

\begin{landscape}
    \begin{figure}[]
    	\centering\includegraphics[width=0.9\linewidth]{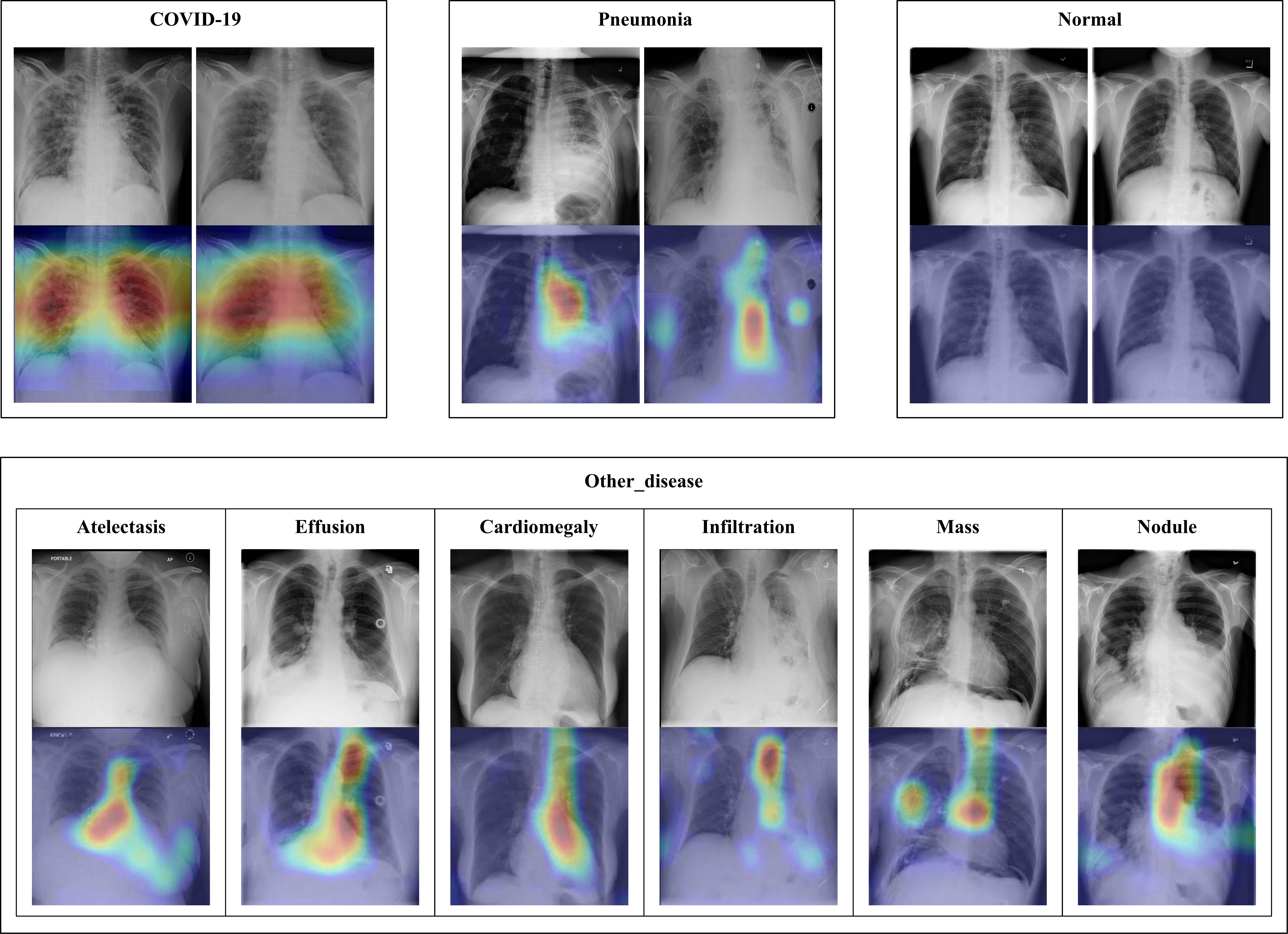}
    	\caption{Visual results obtained by Grad-CAM on different disease classes.}
    	\label{fig:fig3}
    \end{figure}
\end{landscape}

Visualization of the final feature map $\mathcal{A}^i$ depicts the discriminate region of the image, and is obtained by the weighted summation of all the feature maps as 
\begin{equation}
    \mathcal{S}^c \sim= \sum_{i=1}^{K}\omega_i^c\mathcal{A}^i\in \mathbb{R}^{H \times W}, 
    \label{eqn:1}
\end{equation}
where $\omega_i$ controls the importance of individual feature maps depending on the class of interest $c$, and can be measured as the gradient of the $c$th class score with respect to feature maps $A^i$. This is represented  as 
\begin{equation}
    \omega_i^c = \frac{1}{r \times c}\sum_{h=1}^H\sum_{w=1}^W\frac{\partial y^c}{\partial A^k_{h, w}},
    \label{eqn:2}
\end{equation}
and measures the linear effect of the $(h, w)$th pixel in the $k$th feature map on the $c$th class score $y^c$. Finally $\mathcal{S}^c_{Grad-CAM}$ is computed as $ReLU(\mathcal{S}^c)$, as we are  interested only in the features that have a positive influence on the class of interest. Fig. \ref{fig:fig3} illustrates the class activation maps for sample patients from the validation dataset, corresponding to the four classes $normal$, $other\_disease$, $pneumonia$ and $Covid-19$. The red regions in the figure represent areas where the network focuses the most, {\it ie.}, paid more attention during the classification; the blue regions, on the other hand, are the least important regions.
As observed from Fig. \ref{fig:fig3}, in the COVID-19 X-Ray images network focused on the ground glass opacity which is considered as the most prevalent clinically observed pathology for COVID-induced pneumonia  \cite{doi:10.1148/radiol.2020200843}. For the non-COVID Pneumonia cases, on the other hand, the network highlighted typical lung inflammation indicative of pneumonia. In Normal cases no highlighted regions were observed. For the case {\it other\_diseases} the model could correctly focus on the relevant abnormality \cite{meng20}.

\section{Conclusions}

This paper presented Domain Extension Transfer Learning (DETL) for alternative screening of COVID-19 by determining characteristic features from chest X-Ray images. In order to get an idea about the COVID-19 detection transparency, we employed the concept of Gradient Class Activation Map (Grad-CAM) for detecting the regions where the model paid more attention during the classification. This was found to strongly correlate with clinical findings, as validated by experts.

The research is very timely and necessary at the worldwide level. If successful, it will act as assistive intelligence to medical practitioners, for effective handling of the gravity of this pandemic. The initial results show promise, with the possibility of replication on bigger and more diverse data sets.  
\nocite{basu2020deep}
\bibliographystyle{ieeetr} 
\bibliography{covid}

\end{document}